# Water dispersible CoFe$_2$O$_4$ nanoparticles with improved colloidal stability for biomedical applications

Sandeep Munjal[1], Neeraj Khare*,[1], Chetan Nehate[2] and Veena Koul[2]

[1] Department of Physics, Indian Institute of Technology Delhi, Hauz Khas, New Delhi-110016, India.

[2] Centre for Biomedical Engineering, Indian Institute of Technology Delhi, Hauz Khas, New Delhi-110016, India.

## Abstract

Single phase cobalt ferrite (CoFe$_2$O$_4$, CFO) nanoparticles of a controlled size (~ 6 nm) exhibiting superparamagnetic properties have been synthesized by hydrothermal technique using oleic acid (OA) as surfactant. The oleic acid coated CFO nanoparticles are stable in non-polar organic media, such as hexane but are not well dispersible in water. The surface of these snanoparticles has been further modified by citric acid using ligand exchange process, which makes CFO nanoparticles more stable colloidal solution in water. Citric acid coated CFO nanoparticles exhibits high dispersibility in water, high zeta potential, very low coercivity and moderate saturation magnetization. Biocompatibility of these CFO nanoparticles is demonstrated through cytotoxicity test in L929 cell line.

1. **Introduction**

In recent years magnetic nanoparticles have been extensively explored for their potentiality in many biomedical applications such as for targeted drug delivery [1], as contrast enhancement agents in magnetic resonance imaging (MRI) [2], and in hyperthermia treatments as heat mediators [3].The main advantage of magnetic nanoparticles for biomedical applications is its larger surface area for easy ligand attachment, better tissue diffusion and reduced dipole-dipole interaction. The magnetic properties of the nanoparticles can be tuned by controlling its size [4], composition [5], shape [6] and strain/defects [7]. By carefully reducing its size below to a critical diameter, the magnetic nanoparticles can be turned to superparamagnetic nanoparticles.

Iron oxide magnetic nanoparticles such as $Fe_3O_4$ [2] and $\gamma - Fe_2O_3$ [8] have been widely explored for biomedical applications. The saturation magnetization and hysteresis losses of these iron oxide nanoparticles are small compared to pure metals (Co, Fe, or CoFe etc.), but the metallic nanoparticles are highly toxic and very sensitive to oxidation and hence are not useful for biomedical applications. Another alternative can be spinel ferrites such as $MFe_2O_4$ (M $\cong$ Co, Mn, Ni) [9, 10, 11]. Among these ferrites, $CoFe_2O_4$ is interesting due to its large curie temperature, high effective anisotropy and moderate saturation magnetization [12]. CFO has an inverse spinel structure with general formula $AB_2O_4$ (A = Fe and B = Co, Fe) where half of the $Fe^{3+}$ occupies the octahedral sites and the other half $Fe^{3+}$ occupies the tetrahedral sites whereas all the $Co^{2+}$ occupy the octahedral sites.

For biomedical applications the magnetic nanoparticles should be of small sizes with narrow size distribution. These nanoparticles should be coated with some organic or inorganic material which ensure their biocompatibility, nontoxicity and colloidal stability in biophase. Several techniques such as microemulsion [13], coprecipitation [14], ball milling [15], sol−gel [16], thermal decomposition [17], sonochemical [18] and electrosynthesis [9] method have been employed for the synthesis of magnetic nanoparticles but all these synthesis methods often produce larger size nanoparticles with wide particle size distribution.

In the present work, we have synthesized uniform size (~6nm) $CoFe_2O_4$ magnetic nanoparticles using hydrothermal techniques with oleic acid as surfactant. These oleic acid coated CFO nanoparticles are not dispersible in water and in order to make these nanoparticles water dispersible, the surface of these oleic acid coated nanoparticles was modified with citric acid using ligand exchange method. It is found that these citric acid coated $CoFe_2O_4$ nanoparticles makes a good colloidal solution in water in a wide range of pH. The biocompatibility of citric acid coated CFO nanoparticles was studied with mouse fibroblast L929 cells lines, using a MTT cytotoxicity assay.

## 2. Experimental

CFO nanoparticles were synthesized by hydrothermal method [7], using ferric and cobalt nitrate as precursors and oleic acid as surfactant. A solution of 2 mmol of ferric nitrate and 1 mmol of cobalt nitrate in 20 ml water was added with a 10 mmol NaOH solution, ethanol and 12 ml oleic acid. The resultant solution was mixed thoroughly using a magnetic stirrer and was put into a Teflon lined stainless steel autoclave. The autoclave was placed into a preheated oven at 180º C for 16 hours. After cooling, the particles were washed several times in hexane and ethanol. A permanent magnet was used for the separation of the nanoparticles from the liquid. These nanoparticles is named as OA-CFO.

For synthesizing citric acid coated CFO nanoparticles from the oleic acid coated CFO nanoparticles, ligand exchange method was used. The OA-CFO nanoparticles were kept in a solution of toluene, citric acid and dimethyl sulfoxide (DMSO) and stirred thoroughly for 30 hours. Modified nanoparticles were collected, washed in ethanol and dried at 60 ºC. These citric acid coated nanoparticles is named as CA-CFO in the subsequent discussion.

Structural properties of CFO samples were investigated using Rigaku Ultima IV X-ray diffractometer (XRD) equipped with Cu Kα (λ =1.542 Å) radiation source and the morphology of the samples were characterized by using a JEOL JEM-2200-FS Transmission electron microscope (TEM). Magnetic measurements of the CFO nanoparticles were performed at room temperature using Quantum Design Evercool-II Physical property measurement system, in the magnetic field range of -4 to 4 Tesla. For the magnetic measurements samples were prepared by adding a known amount of CFO nanoparticles in DI water and then drop casted onto a glass substrate of dimensions 0.4cm × 0.4 cm. FTIR studies of OA-CFO and CA-CFO nanoparticles were carried out using Thermo Scientific™ Nicolet™ iS™ 50 FT-IR Spectrometer. Zeta potential (ζ) and hydro dynamic diameter of CA-CFO nanoparticles were studied using malvern zetasizer nano zs90.

In vitro cell viability studies of CA-CFO nanoparticles were carried out with mouse fibroblast L929 cells lines, using a 3-(4 5-dimethylthiazol-2-yl)-2 5-diphenyltetrazolium bromide (MTT) cytotoxicity assay. The MTT cytotoxicity assay is a colorimetric assay which measures the cellular metabolic activity based on mitochondrial NADPH dependent dehydrogenase enzymes [19]. These enzymes reduce the MTT dye to form formazan crystals in viable cells. DMSO dissolves these crystals to give a purple colored solution, which can then be quantified spectrophotometrically at 540 nm, using microplate spectrophotometer (PowerWave XS2, BioTek Instruments, USA).

### 3. Results and discussion

XRD patterns of OA-CFO and CA-CFO nanoparticles are shown in Fig. 1. The observed peaks at 2θ = 30.12º, 35.44º, 43.08º, 53.52º, 57.04º and 62.58º corresponds to (220), (311), (400), (422), (511) and (440) planes of $CoFe_2O_4$ (JCPDS No. 22-1086). This confirms the formation of single phase cubic spinel structure of $CoFe_2O_4$ nanoparticles. The XRD patterns of OA-CFO and CA-CFO nanoparticles are similar because these nanoparticles have same crystalline core of $CoFe_2O_4$.

Scherer formula is used to determine the average crystallite size for CFO nanoparticles, which is given as [20];

$$t = \frac{0.9\lambda}{\beta \cos\theta} \quad (1)$$

where $\beta$ represents the full width at half maximum of the XRD peak, $\theta$ is the Bragg's angle, $\lambda$ (1.542 Å) is the wavelength of X-ray, and t is the average crystallite size. The average crystallite sizes for OA-CFO and CA-CFO nanoparticles are found as ~5.9 nm and 5.4 nm respectively.

TEM images of OA-CFO nanoparticles and CA-CFO nanoparticles are shown in Fig. 2. The histogram of the particle size distribution are also shown as inset in Fig. 2a and 2b. The size distribution of the CFO nanoparticles reveals that the maximum number of the particles has a diameter in the size range of 4 nm to 7 nm with a log normal peak appearing at 5.8 and 5.6 nm for OA-CFO and CA-CFO nanoparticles. These results are in good agreement with the crystallite size obtained from XRD analysis.

Fig. 3 shows the field dependence of magnetization for OA-CFO and CA-CFO nanoparticles at room temperature. OA-CFO nanoparticles are found to have a slightly high saturation magnetization as compared to CA-CFO nanoparticle. For OA-CFO nanoparticles the value of the saturation magnetization ($M_s$) and coercivity ($H_c$) are found as ~47 emu/gm and ~11 Oe respectively, and for CA-CFO nanoparticles, the values of the $M_s$ and $H_c$ are found as ~42 emu/gm and ~13 Oe respectively. The small values of coercivity of CFO nanoparticles indicates that these nanoparticles are near the superparamagnetic limit which is the ideal regime for several biomedical applications.

The lower value of saturation magnetization ($M_s$) of these CFO nanoparticles compared to the Ms value of bulk cobalt ferrite (~ 90 emu/gm) can be attributed to a "nanoscale size effect" according to which the magnetic moments present near the surface of the nanoparticles behaves differently from that present in the core of the particles and a much higher spin disorder is present on the surface of the particles which leads to the reduction of $M_s$ of CFO nanoparticles [21].

FTIR spectrum of OA-CFO and CA-CFO nanoparticles are shown in Fig. 4. The difference in FTIR spectra of OA-CFO and CA-CFO nanoparticles is due to the presence of different coating on CFO for both of these two samples. The presence of oleic acid on as synthesized OA-CFO nanoparticles was confirmed by two $CH_3$ stretching at 2920 $cm^{-1}$ and 2850 $cm^{-1}$ present in FTIR spectra of the sample. The two bands appears near 1538 and 1410 $cm^{-1}$, which are characteristic bands of the asymmetric and the symmetric stretch of (COO). It is evident that oleic acid was chemisorbed onto the surface of CFO nanoparticles via its carboxylate group [22].

The CA-CFO nanoparticles shows three strong absorption peaks at 3275, 1575, and 1405 $cm^{-1}$ corresponding to the stretching band of hydroxyl group (-OH), antisymmetric $v_{as}$ (COO) and symmetric $v_s$ (COO) stretching band of the carboxyl group, respectively [23]. This confirms that the surface of the CA-CFO nanoparticles was covered with carboxylate species of citric acid. An intense peak at ∼590 $cm^{-1}$ is observed, which is attributed to the stretching of the metal ion at the tetrahedral A-site, $M_A \leftrightarrow O$ [24].

The CA-CFO nanoparticles are well dispersible in water in a wide range of pH. The hydrodynamic diameter ($D_H$) determined by Dynamic Light Scattering (DLS) and zeta potential ($\zeta$) values for CA-CFO nanoparticles at different pH values ranging from 2.2 to 10.8, are shown in Fig. 5. As the $\zeta$ is related to the surface charge present on the nanoparticles, the large/small value of |ζ| indicates the more/less electrostatic repulsion between the nanoparticles. In the case of magnetic nanoparticles this electrostatic repulsion opposes the magnetic attraction acting between the nanoparticles. Initially increasing the pH from 2.2 the |ζ| approaches toward "0" and at pH value ~3.5, |ζ| is minimum. At this point the magnetic attraction exceeds the electrostatic repulsion which leads to more agglomeration of CA-CFO nanoparticles. Due to this agglomeration the $D_H$ shows its maximum value at pH ~3.5. At higher pH values (> 3.5) the |ζ| starts increasing, which increases the electrostatic repulsion between the nanoparticles that leads to a decrease in $D_H$. At pH ~7, ζ is sufficiently negative (-22.3 mV), which indicates that negative charges are present on the surface of CA-CFO nanoparticles in a larger amount. Smaller value of $D_H$ for CA-CFO nanoparticles indicates that good electrostatic repulsive forces are acting between the particles that opposes the agglomeration of these nanoparticles and increases the dispersion and colloidal stability in water. All these observations suggests that the surface modification of OA-CFO nanoparticles by citric acid allows us to obtain a good dispersion in water with more colloidal stability.

The OA-CFO nanoparticles are not dispersible in water so these nanoparticle cannot be used in biomedical applications but the high colloidal stability of CA-CFO nanoparticles at neutral pH values make the CA-CFO magnetic nanoparticles a suitable candidate for biomedical applications. The biocompatibility of citric acid coated CFO nanoparticles was studied on the L929 (mouse fibroblast) cells line, using an MTT cytotoxicity assay. Cells were seeded at a density of $10 \times 10^3$ cells per well in 96 well plate. The plate was incubated at 37 °C for 24 h in a $CO_2$ incubator. CA-CFO nanoparticles suspension was added to each well so that final concentration range was from 100 µg/ml to 1000 µg/ml and incubated for another 24 h. The media was replaced with fresh media and 10 µL of 5% MTT solution, filtered through sterile 0.22 µm filter was added to each well and incubated for 4 h. Cell viability was calculated relative to negative control (phosphate buffered saline) and a positive control (1% Triton X-100) using the following relation:

$$\text{Cell Viability (\%)} = \frac{\text{Sample}_{540nm} - \text{Positive Control}_{540nm}}{\text{Negative Control}_{540nm} - \text{Positive Control}_{540nm}} \times 100$$

Fig. 6 shows the surviving fraction of L929 cells incubated during 24 hours with different concentrations of CA-CFO nanoparticles and evaluated by the MTT assay. The synthesized CA-CFO nanoparticles were found to biocompatible at relatively high concentrations (100 μg/ml to 1000 μg/ml), with cell viability almost 100%, i.e. the survival fraction of treated cells is similar to the controls cell under the experimental conditions. This high cell viability of CA-CFO nanoparticles is attributed to the coating of citric acid on these nanoparticles through ligand exchange method. As the zeta potential of CA-CFO nanoparticles was found to be large (-18.8 mV), these CA-CFO nanoparticles will have long circulatory effect in biophase.

## 4. Conclusion

We have successfully synthesized oleic acid capped, uniform size (~6 nm) CFO nanoparticles by hydrothermal method and demonstrated the conversion of these nanoparticles to a citric acid (CA) coated CFO nanoparticles using ligand exchange method. The CA-CFO nanoparticles exhibits very small coercivity, moderate saturation magnetization and a high degree of dispersibility in water. At pH ~ 7, zeta potential of CA-CFO nanoparticles was higher (~22 mV), which gives CA-CFO nanoparticles colloidal stability in water. Cell viability assay of CA-CFO nanoparticles on L929 cell line showed no cytotoxic effect even after 24 h of incubation period at relatively higher concentration of CA-CFO nanoparticles, which establishes the potentiality of the functionalized CA-CFO nanoparticles for biomedical applications.

**Acknowledgments**

The financial support from DeitY (Government of India) is gratefully acknowledged. One of us (SM) is also thankful to Council of Scientific and Industrial Research (CSIR), New Delhi for senior research fellowship (SRF) Grant.

**Figure Captions**

Fig. 1  X-ray diffraction patterns of OA-CFO and CA-CFO nanoparticles.

Fig. 2.  TEM images of (a) OA-CFO nanoparticles and (b) CA-CFO nanoparticles. The inset in the figure shows distribution of particle size and its log normal distribution fit.

Fig. 3.  Magnetic hysteresis loops for OA-CFO and CA-CFO nanoparticles at room temperature.

Fig. 4.  FTIR spectra of OA-CFO and CA-CFO nanoparticles. Arrows marked OA-CFO spectra corresponds to 2990 cm$^{-1}$, 2850 cm$^{-1}$, 1538 cm$^{-1}$, 1410 cm$^{-1}$ and 590 cm$^{-1}$ (from left to right) whereas the arrows marked to CA-CFO spectra corresponds to 3275 cm$^{-1}$, 1575 cm$^{-1}$, 1405 cm$^{-1}$ and 590 cm$^{-1}$ (from left to right).

Fig. 5.  Zeta potential and hydrodynamic diameter as a function of pH for CA-CFO nanoparticles.

Fig. 6.  Cytotoxicity profiles of CA-CFO nanoparticles for 24 h on L929 cell line at different concentrations (100, 400, 600, 800 and 1000 μg/mL)

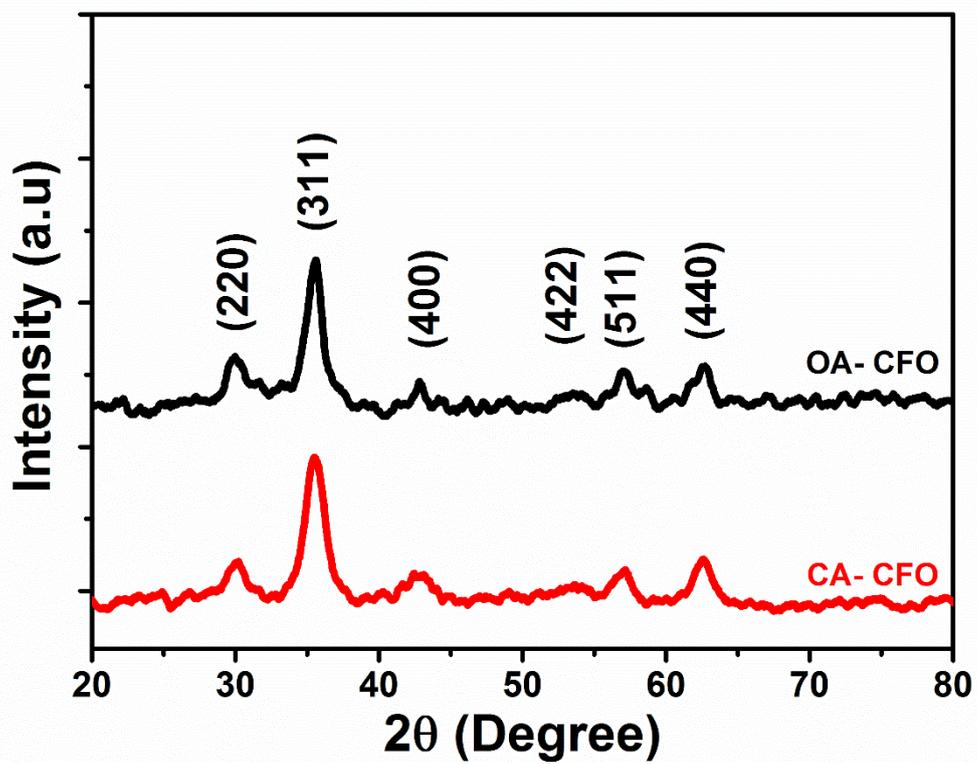

Fig. 1  X-ray diffraction patterns of OA-CFO and CA-CFO nanoparticles.

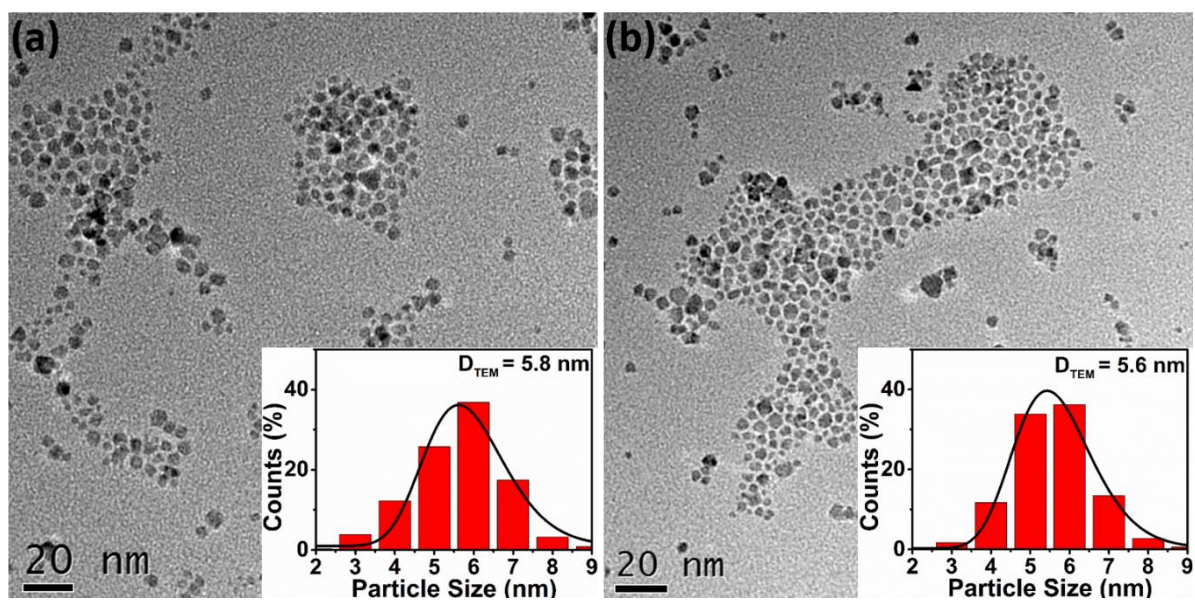

Fig. 2. TEM images of (a) OA-CFO nanoparticles and (b) CA-CFO nanoparticles. The inset in the figure shows distribution of particle size and its log normal distribution fit.

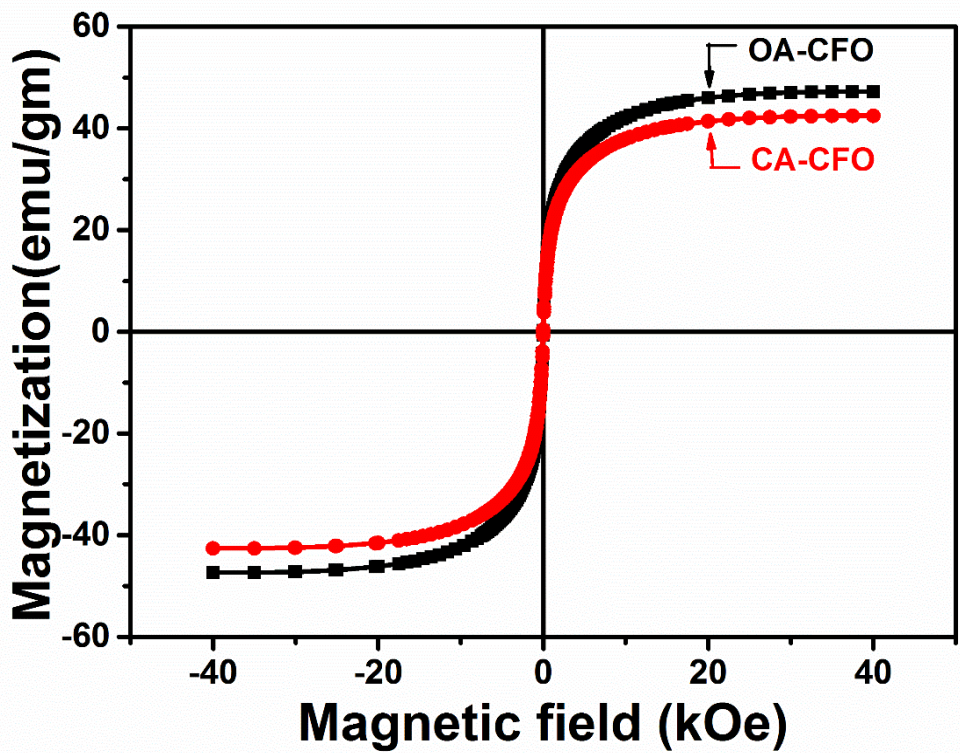

Fig. 3. Magnetic hysteresis loops for OA-CFO and CA-CFO nanoparticles at room temperature.

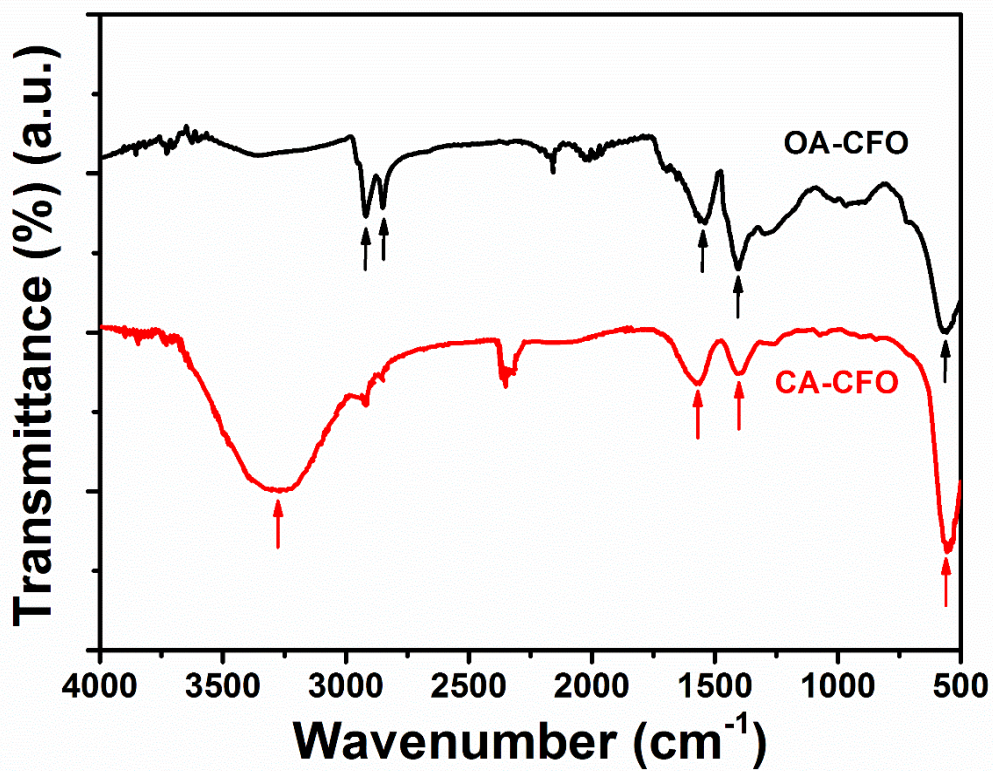

Fig. 4. FTIR spectra of OA-CFO and CA-CFO nanoparticles. Arrows marked OA-CFO spectra corresponds to 2990 cm$^{-1}$, 2850 cm$^{-1}$, 1538 cm$^{-1}$, 1410 cm$^{-1}$ and 590 cm$^{-1}$ (from left to right) whereas the arrows marked to CA-CFO spectra corresponds to 3275 cm$^{-1}$, 1575 cm$^{-1}$, 1405 cm$^{-1}$ and 590 cm$^{-1}$ (from left to right).

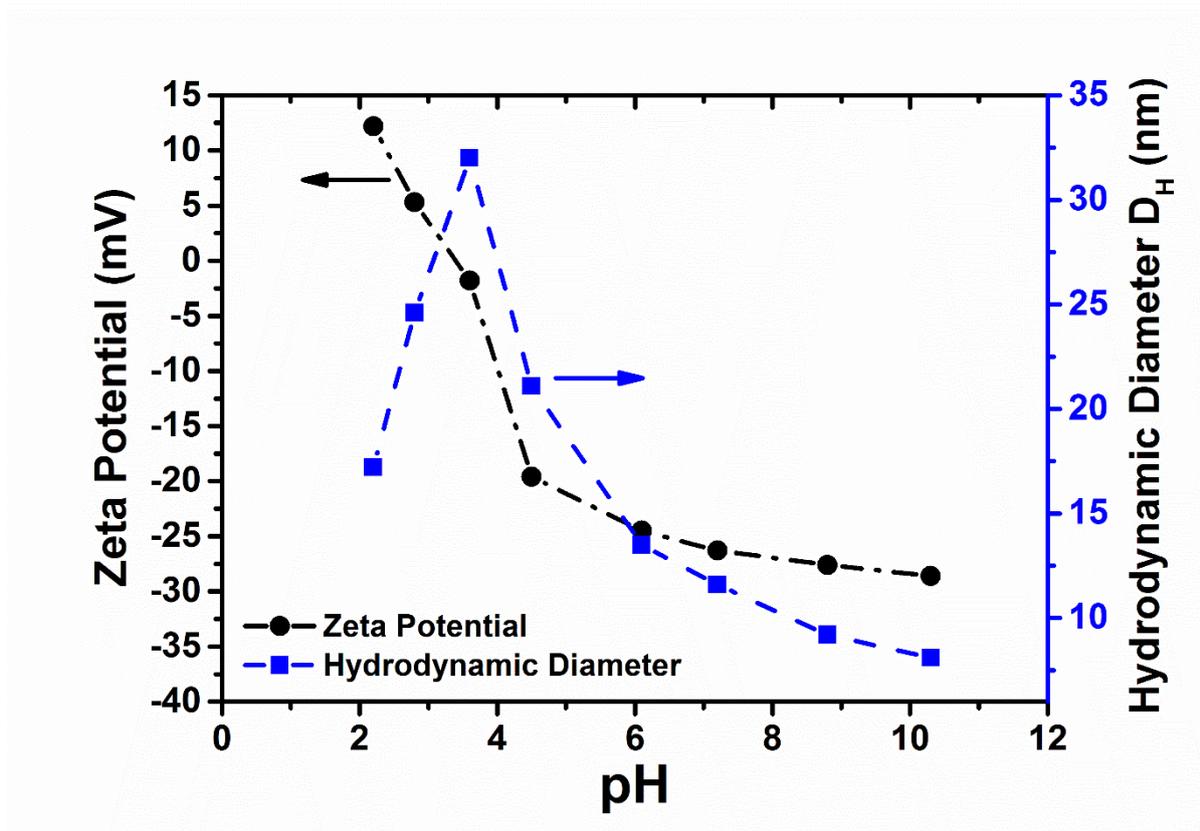

Fig. 5. Zeta potential and hydrodynamic diameter as a function of pH for CA-CFO nanoparticles.

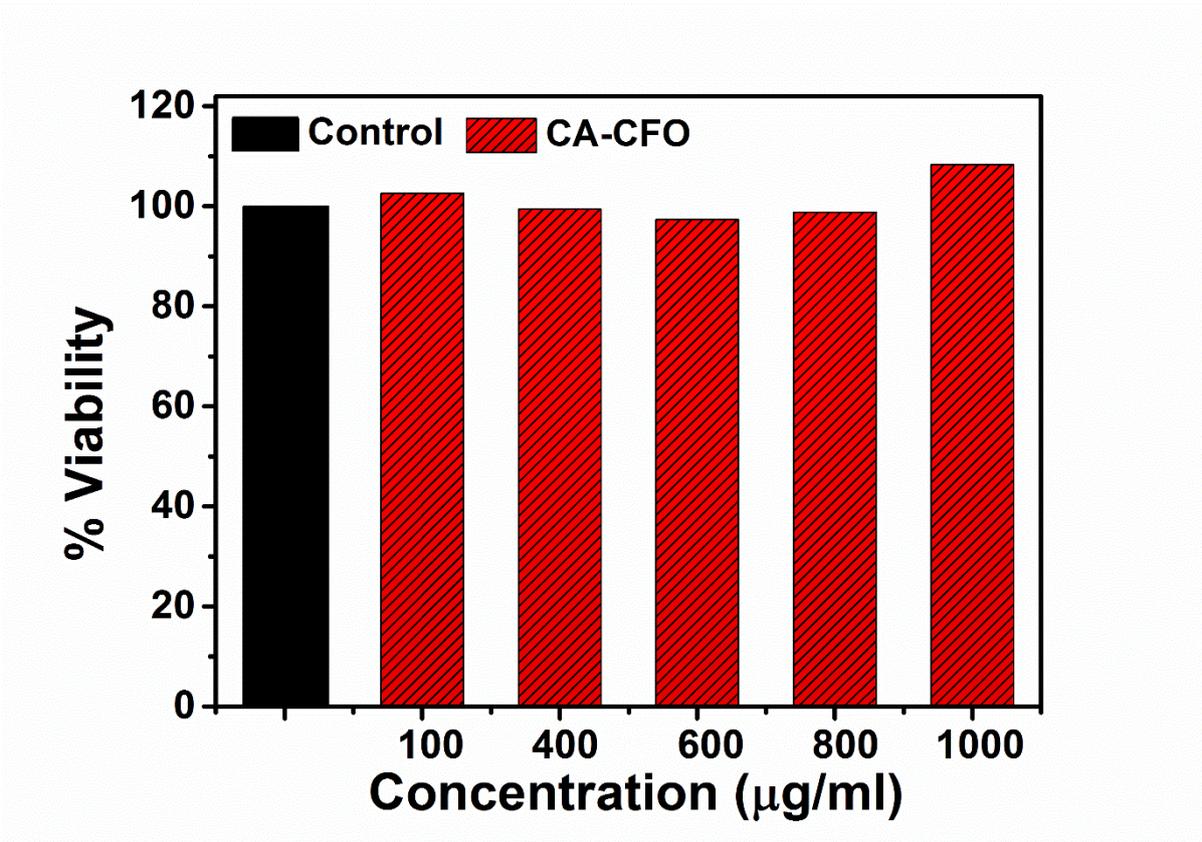

Fig. 6. Cytotoxicity profiles of CA-CFO nanoparticles for 24 h on L929 cell line at different concentrations (100, 400, 600, 800 and 1000 μg/mL).